\documentclass{IEEEcsmag}
\IEEEoverridecommandlockouts
\usepackage{flushend}
\usepackage{caption}
\usepackage{subfigure}
\usepackage{subcaption}
\usepackage{graphicx}
\usepackage[table, svgnames, dvipsnames]{xcolor}
\usepackage{booktabs}
\usepackage{comment}
\usepackage{cite}
\usepackage{amsmath,amssymb,amsfonts}
\usepackage{algorithmic}
\usepackage{textcomp,balance,url}
\usepackage{multirow}
\usepackage{graphicx,array}
\usepackage{ragged2e}
\newcolumntype{C}[1]{>{\centering\let\newline\\\arraybackslash\parindent=0pt\hspace{0pt}}m{#1}}
\newcolumntype{L}[1]{>{\justifying\let\newline\\\arraybackslash\parindent=0pt\hspace{0pt}}m{#1}}
\newcolumntype{T}[1]{>{\raggedright\let\newline\\\arraybackslash\parindent=0pt\hspace{0pt}}m{#1}}
\usepackage[most]{tcolorbox} % for a flexible, professional box
\definecolor{chestnut}{rgb}{0.97, 0.51, 0.47}
\def\BibTeX{{\rm B\kern-.05em{\sc i\kern-.025em b}\kern-.08em
    T\kern-.1667em\lower.7ex\hbox{E}\kern-.125emX}}
\usepackage{amsmath,amsfonts}
\usepackage{subfigure}
\usepackage{adjustbox}
\usepackage{comment}
\usepackage{soul}
\usepackage{textcomp}
\usepackage{siunitx}
\usepackage{booktabs}
\usepackage{stfloats}
\usepackage{xcolor}
\usepackage[most]{tcolorbox}
\usepackage{float}
\definecolor{lightboxblue}{RGB}{200,234,243}
\definecolor{lightboxgreen}{RGB}{222,241,204}
\definecolor{lightboxpink}{RGB}{252,235,238}
\usepackage{url}
\usepackage{verbatim}
\let\labelindent\relax
\usepackage{enumitem}
\usepackage{multirow}
\definecolor{IEEEblue}{RGB}{0,102,153}

\usepackage{ragged2e}
\usepackage{cite}
\usepackage{float}

\usepackage[table]{xcolor}
\usepackage{tikz}
\definecolor{stageblue}{RGB}{0,45,120}
\definecolor{stagegreen}{RGB}{0,120,35}
\definecolor{stageorange}{RGB}{230,95,0}

\graphicspath{{./images/}}
\jmonth{September}
\jname{IEEE Internet Computing}
\jtitle{From Pixels to Semantics: Edge AI for UAV-Based Critical Infrastructure Inspection}
\pubyear{2026}

\begin{document}
\sptitle{IEEE Internet Computing} 
\title{From Pixels to Semantics: Edge AI for UAV-Based Critical Infrastructure Inspection}
\author{Reza Farahani *}
\affil{Distributed Systems Group, TU Wien, Austria}
\author{Naser Hossein Motlagh}
\affil{University of Helsinki, Finland}
\author{Zoha Azimi}
\affil{University of Klagenfurt, Austria}
\author{Christian Timmerer}
\affil{University of Klagenfurt, Austria}
\author{{Lorenzo Carnevale}}
\affil{University of Messina, Italy}
\author{{Sasu Tarkoma}}
\affil{University of Helsinki, Finland}
\author{Schahram Dustdar}
\affil{Distributed Systems Group, TU Wien, Austria and ICREA, Barcelona, Spain\\[4pt]
\small $^{\ast}$ Corresponding author: r.farahani@dsg.tuwien.ac.at
}
\markboth{IEEE Internet Computing}{IEEE Internet Computing}
\begin{abstract} \justifying
Critical infrastructure assets such as bridges, tunnels, dams, and power line networks require timely and scalable inspection. While conventional manual inspection remains costly and hazardous, unmanned aerial vehicle (UAV)-based inspection has emerged as an efficient alternative for monitoring difficult-to-access structures. Existing UAV inspection pipelines have evolved from cloud-centric offline processing toward edge-based perception using lightweight object detectors such as YOLO for real-time defect localization. This article explores the transition toward fully edge-native semantic inspection powered by lightweight vision language models (VLMs), where UAVs move beyond object detection toward contextual structural understanding. It categorizes existing UAV inspection architectures, identifies their key system challenges and architectural requirements, and experimentally assesses the feasibility of semantic edge intelligence on NVIDIA Jetson UAV-class hardware using the COCO-Bridge dataset. The evaluation integrates a fine-tuned YOLO-26M for object localization and a lightweight SmolVLM-256 for semantic reasoning. Finally, it outlines future directions toward agentic, autonomous, trustworthy, and collaborative semantic UAV inspection across the edge-cloud continuum.
\end{abstract}
\maketitle
\section{Introduction}
\label{sec:Introduction }
\begin{figure*}
    \centering
    \includegraphics[width=.9\linewidth]{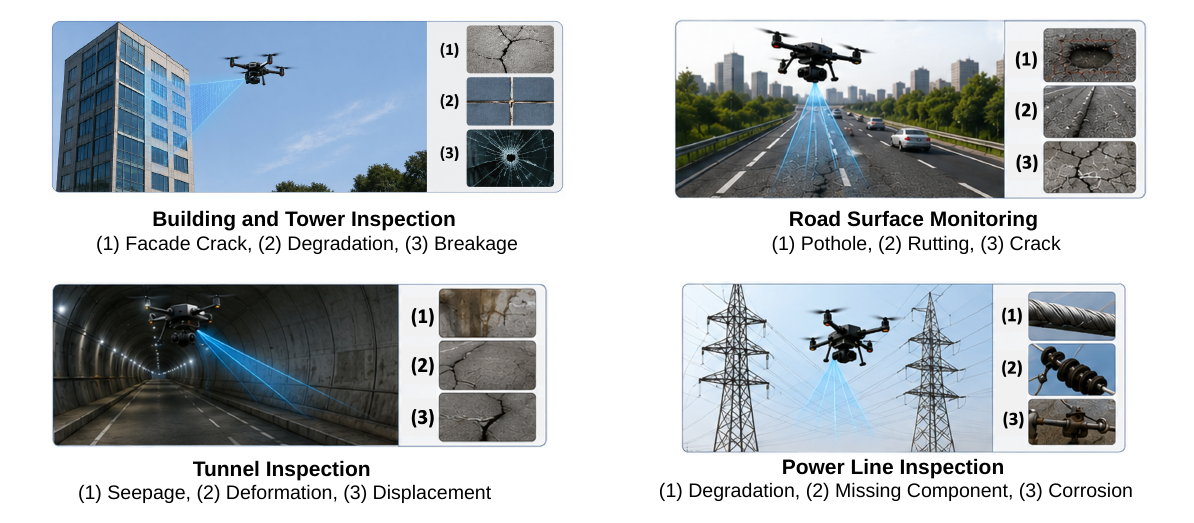}
    \caption{Representative UAV-based infrastructure inspection scenarios and related structural anomalies.}
    \label{fig:example}
\end{figure*}

%%%
The aging and growing scale of critical infrastructure systems demand inspection solutions that are faster, safer, and more scalable than conventional manual procedures. Recent advances in unmanned aerial vehicles (UAVs), commonly known as drones, high-resolution sensing, and artificial intelligence (AI) are rapidly transforming infrastructure monitoring by enabling automated visual inspection across large-scale and difficult-to-access environments. As shown in Fig.~\ref{fig:example}, UAV-based inspection is increasingly applied across diverse infrastructure domains, e.g., facade and tower assessment, road surface monitoring, tunnel inspection, and power-line analysis. These scenarios involve heterogeneous structural anomalies, such as cracks, corrosion, seepage, displacement, rutting, and component degradation that require accurate visual assessment under dynamic operational conditions. 

The combination of autonomous aerial mobility and modern imaging platforms enables UAVs to capture detailed visual information with significantly higher flexibility and coverage than traditional inspection approaches. However, continuous UAV missions also generate massive streams of high-resolution visual data, while only a small fraction of captured frames typically contains inspection-relevant information. Efficiently filtering redundant data and extracting and interpreting relevant visual evidence, therefore, remain key challenges for scalable, real-time inspection pipelines. Rather than serving solely as airborne sensing platforms, recent UAVs incorporate onboard perception and AI inference capabilities to process inspection data during flight. This shift toward onboard intelligence is changing how inspection data are processed, transmitted, and interpreted across edge-cloud environments. 

This article analyzes the evolution of UAV-based infrastructure inspection from cloud-centric visual analysis to onboard edge perception and fully edge-native semantic inspection. Beyond existing paradigms, we investigate the feasibility of deploying lightweight semantic edge intelligence directly on UAV-class hardware through experiments on an NVIDIA Jetson platform using a real-world bridge inspection dataset.” Specifically, we fine-tune and evaluate a lightweight YOLO detector for structural component localization together with a compact vision language model (VLM) for contextual defect interpretation and semantic inspection reporting. Finally, we identify key architectural requirements, operational trade-offs, and outline future directions for intelligent UAV inspection systems.

\section{From Cloud-Centric Inspection to Semantic Edge Intelligence}\label{sec:BR:RelatedWork}
\begin{figure*}[!t]
\centering
  \subfigure[Cloud-centric inspection.]
  {\includegraphics[width=0.3\textwidth]{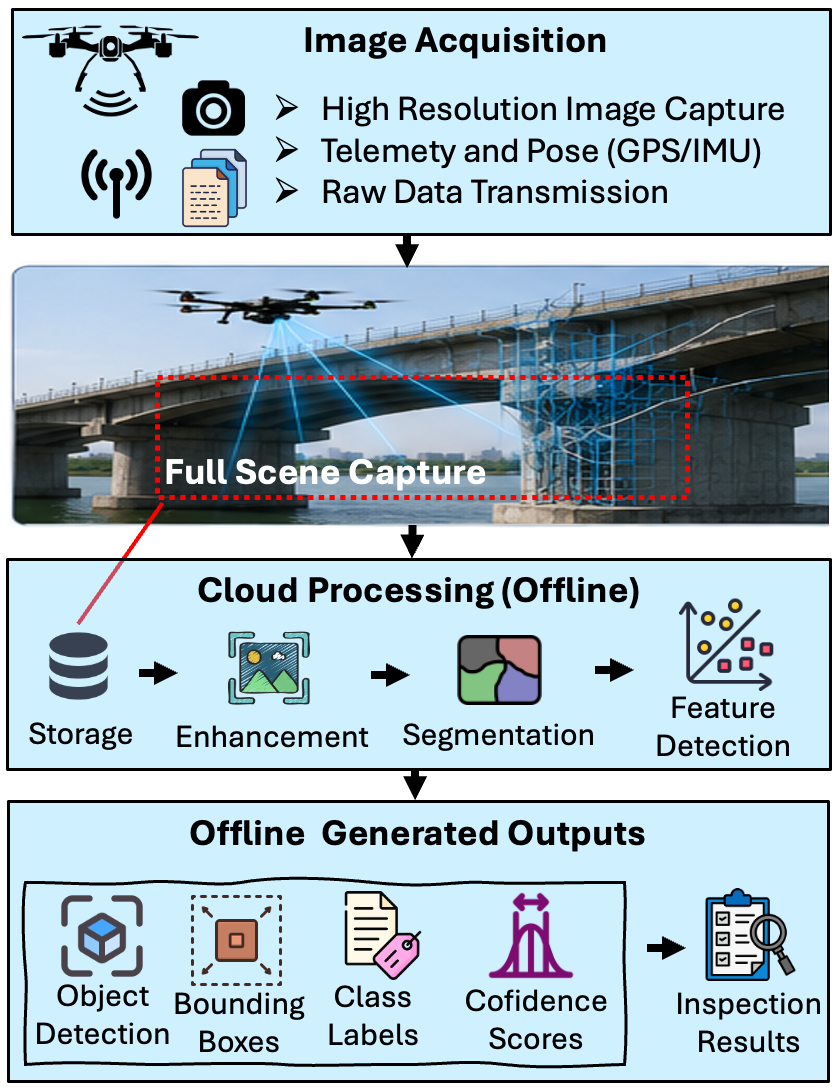}
  \label{fig:Fig1-a}}
  \subfigure[Onboard edge-based inspection.]{\includegraphics[width=0.3\textwidth]{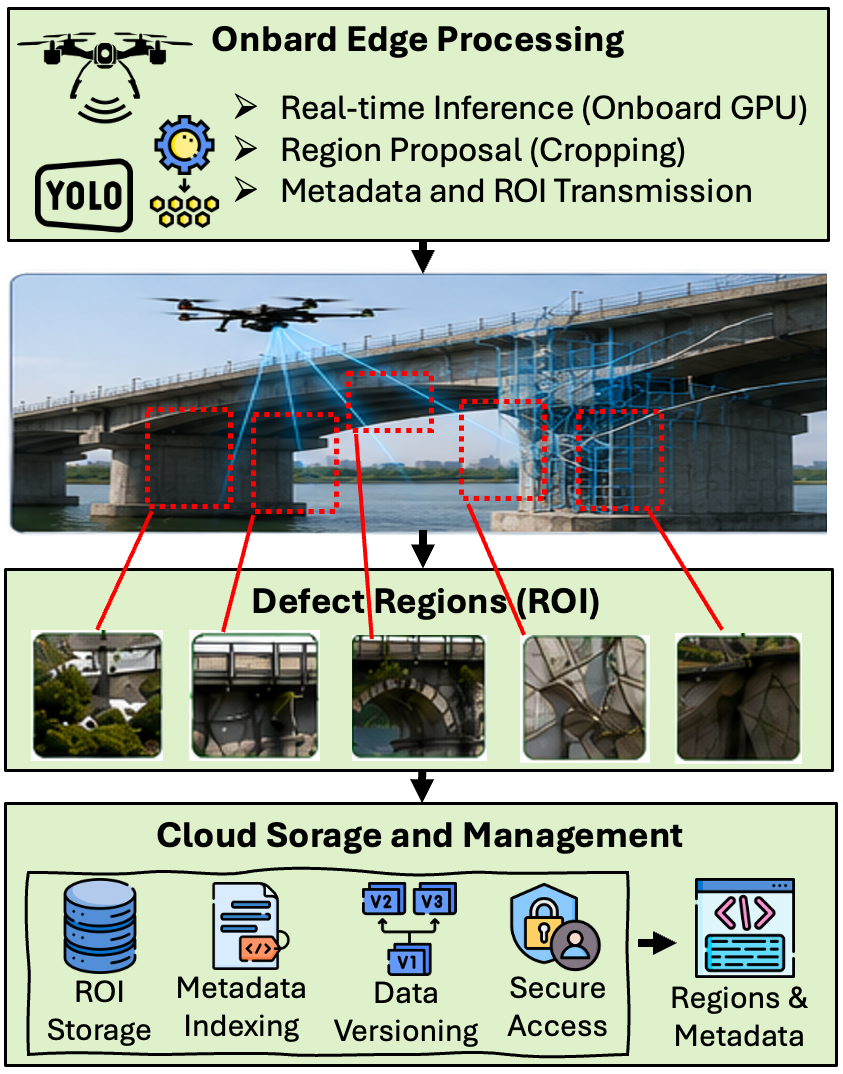}
  \label{fig:Fig1-b}}
  \subfigure[Edge-native semantic inspection.]{\includegraphics[width=0.3\textwidth]{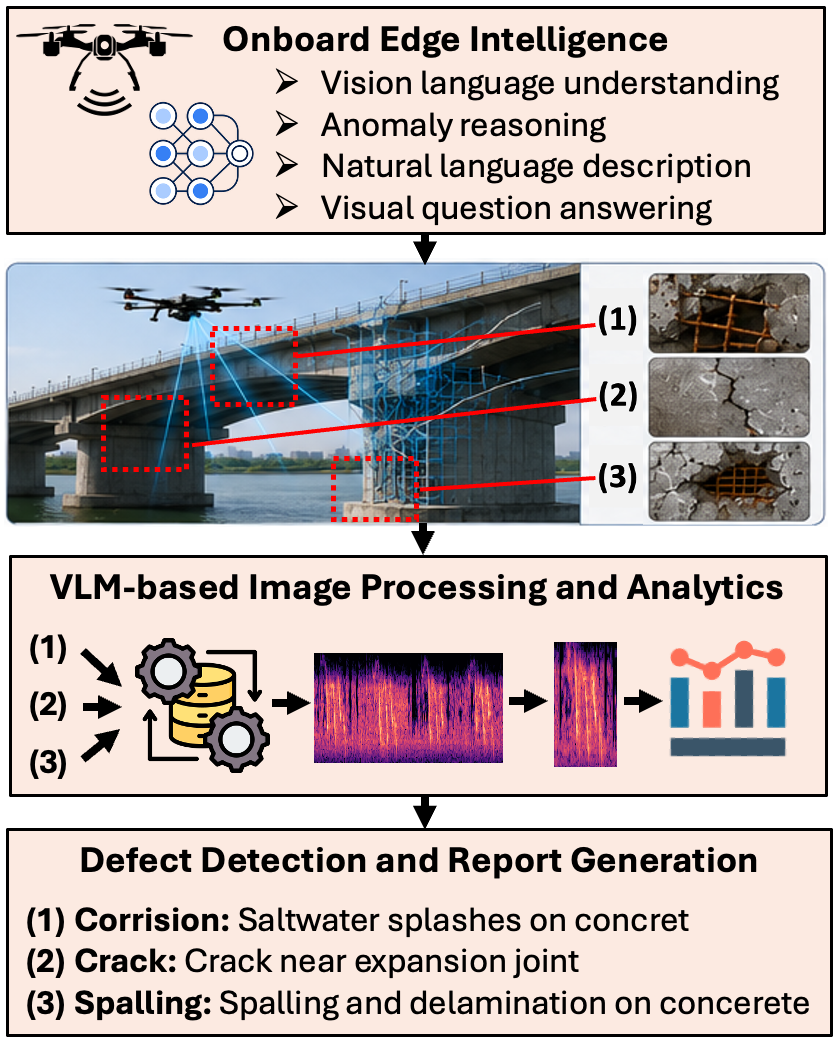}
  \label{fig:Fig1-c}}
\caption{Comparison of UAV infrastructure inspection paradigms according to AI placement and inspection understanding, progressing from cloud-centric visual analysis to edge-native semantic intelligence.}
\label{fig:evolution}
\end{figure*}

Fig.~\ref{fig:evolution} categorizes UAV-based infrastructure inspection pipelines according to the placement and capabilities of AI inference within the inspection workflow. These paradigms differ not only in \emph{where inference is performed}, but also in the \emph{level of semantic information} extracted, ranging from raw visual data acquisition to semantically enriched inspection findings.
%%%

\textbf{Cloud-centric visual inspection}, shown in Fig.~\ref{fig:Fig1-a}, represents the earliest UAV inspection paradigm with no on-board AI capabilities. The UAV primarily serves as an airborne sensing platform, capturing high-resolution imagery along with telemetry and positioning data (e.g., GPS/IMU). The complete visual stream is transmitted to cloud or ground infrastructure, where offline processing stages such as image enhancement, segmentation, feature extraction, and defect detection are executed. The resulting outputs typically include detected anomalies, bounding boxes, class labels, and inspection reports.
Following this approach, Li et al.~\cite{li2023automatic} proposed a UAV-assisted bridge inspection framework using Faster Region-based Convolutional Neural Network (R-CNN) for crack detection, where all captured imagery is processed offline after the inspection mission. Gwon et al.~\cite{gwon2023cnn} addressed image quality assessment for UAV-based bridge inspection using a CNN-based classifier that filters blurred and degraded images before structural analysis.
\begin{tcolorbox}[
    enhanced,
    width=\columnwidth,
    colback=lightboxblue,
    colframe=black,
    coltext=black,
    opacityback=1,
    opacityframe=1,
    boxrule=0.8pt,
    arc=3pt,
    outer arc=3pt,
    left=6pt,
    right=6pt,
    top=4pt,
    bottom=4pt
]
\textbf{State-of-the-art limitations:} While they benefit from powerful remote computation and large-scale model execution, they introduce communication overhead, delayed feedback, and dependency on reliable connectivity due to the continuous transmission of high-resolution visual streams.
\end{tcolorbox}
%%%
\textbf{Onboard edge perception}, illustrated in Fig.~\ref{fig:Fig1-b}, shifts defect detection directly onto UAV-side edge hardware, reducing bandwidth consumption and improving inspection responsiveness. Lightweight deep learning models such as YOLO enable real-time inference during flight, allowing the UAV to perform defect localization and region proposal generation directly on captured frames. Instead of transmitting complete visual streams, it typically extracts and transmits only relevant regions of interest (ROIs), along with metadata, for cloud-side storage and management.
This transition shifts the role of UAVs from passive imaging devices to active edge-perception platforms. 

Following this approach, Ji et al.~\cite{ji2025wdi} proposed a YOLOv8-based framework for steel bridge weld defect detection, optimized using lightweight feature extraction and attention mechanisms. Ruggieri et al.~\cite{ruggieri2025using} enhanced YOLO11 with attention modules for reinforced concrete bridge inspection, improving detection precision under low-resolution conditions while reducing inference latency. Li et al.~\cite{li2025improved} introduced an optimized real-time detection transformer (RT-DETR) architecture for railway obstacle intrusion detection with reduced parameter count and faster inference, while Ding et al.~\cite{ding2024improved} improved RT-DETR for power line inspection by enhancing small-object detection and removing conventional post-processing stages. Rong et al.~\cite{rong2025advanced} employed PL-YOLOv8 for power line inspection, where only vegetation encroachment events exceeding predefined thresholds are transmitted with GPS metadata. 
\begin{tcolorbox}[
    enhanced,
    width=\columnwidth,
    colback=lightboxgreen,
    colframe=black,
    coltext=black,
    opacityback=1,
    opacityframe=1,
    boxrule=0.8pt,
    arc=3pt,
    outer arc=3pt,
    left=6pt,
    right=6pt,
    top=4pt,
    bottom=4pt
]
\textbf{State-of-the-art limitations:}
Compared to cloud-centric pipelines, onboard edge perception reduces unnecessary data transfer and enables faster inspection feedback. However, such approaches remain largely detection-centric, with outputs limited to localized anomalies and predefined defect categories.
\end{tcolorbox}
%%%
\textbf{Semantic edge intelligence with VLMs}, depicted in Fig.~\ref{fig:Fig1-c}, integrates lightweight VLMs directly into the UAV edge pipeline to move beyond object-level detection toward contextual understanding of structural conditions. Through multimodal reasoning during flight, such systems enable anomaly interpretation, natural-language description, and visual question answering. Unlike conventional detection pipelines that output only bounding boxes and confidence scores, VLM-based inspection produces semantically enriched findings that combine localized anomalies with contextual interpretation, enabling actionable infrastructure defect reports.

Existing research already demonstrates the potential of semantic AI for UAV systems. Li et al.~\cite{li2025vlm} employed a VLM-driven framework for power line inspection, where semantic scene understanding and navigation reasoning are performed through VLM and large language model (LLM) integration. 
Zhang et al.~\cite{zhang2026multimodal} combined VLM reasoning with segmentation models for semantic road crack assessment and automated report generation, while Chen et al.~\cite{chen2025bridge} proposed a Contrastive Language-Image Pre-training (CLIP)-based framework for semantic bridge inspection guidance through human-UAV interaction. 
%%%%%
\begin{tcolorbox}[
    enhanced,
    width=\columnwidth,
    colback=lightboxpink,
    colframe=black,
    coltext=black,
    opacityback=1,
    opacityframe=1,
    boxrule=0.8pt,
    arc=3pt,
    outer arc=3pt,
    left=6pt,
    right=6pt,
    top=4pt,
    bottom=4pt
]
\textbf{State-of-the-art limitations:}
These approaches remain largely cloud-centric, while the lightweight and resource-aware deployment of semantic VLM reasoning directly on UAV-side edge hardware remains largely unexplored. This transition introduces new architectural challenges and system requirements, discussed in the next section.
\end{tcolorbox}
%%%
%%%
%%%

\section{System Challenges for Semantic UAV Infrastructure Inspection}
\label{sec:Requirements}
Enabling fully edge-based VLM-driven UAV inspection introduces fundamentally different requirements and challenges compared to cloud-centric or detection-only pipelines. Unlike conventional object detectors that primarily localize predefined defects, semantic edge intelligence requires multimodal reasoning, contextual interpretation, and natural-language generation directly on resource-constrained UAVs. This transition introduces new challenges, summarized in Table~\ref {tab:challenge}.
%%%%%%%%%%%%%
\begin{table*}[t]
\caption{Key requirements, challenges, and optimization techniques for semantic UAV infrastructure inspection.}
\label{tab:challenge}
\centering
\small
\fontsize{7.5pt}{7.5pt}\selectfont
\begin{tabular}{p{3.1cm} p{5.0cm} p{4.9cm}}
\toprule
\textbf{Requirement} & \textbf{Challenge} & \textbf{Optimization Techniques} \\
\midrule

Edge sensing and hardware &
High-resolution semantic inspection under limited battery, memory, thermal, and payload constraints. &
Lightweight VLMs, TensorRT, quantization, pruning, embedded accelerators. \\
\midrule

Low-latency multimodal reasoning &
Low-latency multimodal reasoning during flight-time operation. &
Adaptive inference, selective offloading, PEFT/LoRA, dynamic model scaling. \\
\midrule

Hierarchical semantic processing &
Continuous VLM inference over all frames is computationally inefficient. &
YOLO/RT-DETR cascades, ROI filtering, key-frame triggering. \\
\midrule

Mission-aware operation &
Inference delay increases hover time, propulsion energy, and mission duration. &
Joint flight-inference scheduling, adaptive frame sampling, energy-aware planning. \\
\midrule

Multimodal training data &
Limited infrastructure image-text datasets and semantic annotations. &
Synthetic captioning, weak supervision, instruction tuning. \\
\midrule

Cross-structure generalization &
Defect appearance varies across materials, geometry, and environments. &
Domain adaptation, continual learning, and structure-aware fine-tuning. \\
\midrule

Trustworthy autonomy &
Incorrect semantic decisions may affect mission planning. &
Confidence estimation, XAI, RAG grounding, and human verification. \\
\bottomrule
\end{tabular}
\end{table*}
%%%%%%%%%%%%%

\textbf{Edge Hardware and Sensing.}
Semantic UAV inspection requires integrating high-resolution imaging sensors, embedded accelerators, and energy-efficient communication modules that support real-time multimodal inference during flight. Since structural anomalies such as cracks, corrosion, and spalling often occupy only a small fraction of captured frames, stable high-resolution sensing and precise image acquisition become critical for reliable semantic interpretation. At the same time, UAVs remain constrained by limited battery capacity, onboard memory, thermal envelopes, and payload budgets, making lightweight and resource-aware AI deployment essential.

\textbf{Low-latency Multimodal Reasoning}
Unlike offline cloud processing, semantic edge inspection requires tightly coupled perception and reasoning pipelines operating directly during flight. This introduces strict latency, memory, and energy constraints for executing VLM inference on UAV-side hardware. The system must continuously balance local processing, selective offloading, and communication overhead according to runtime conditions such as battery level, wireless link quality, and computational load. Consequently, model compression, quantization, pruning, and adaptive inference strategies become central requirements for practical deployment.

\textbf{Hierarchical Semantic Processing.}
Continuous semantic reasoning over every captured frame is computationally inefficient and often operationally unnecessary. Practical UAV inspection pipelines thus require hierarchical processing architectures, in which lightweight detectors such as YOLO or real-time detection transformer (RT-DETR)~\cite{li2025improved} first identify candidate regions or key frames, triggering semantic VLM reasoning only when a detailed contextual assessment is required. Such hierarchical pipelines are essential for balancing semantic richness, responsiveness, and onboard energy consumption.

\textbf{Mission-aware Scheduling and Flight Constraints.}
Unlike static edge systems, UAV inspection pipelines operate under continuous mobility and strict flight-time constraints. Inspection quality is thus tightly coupled with flight speed, hover duration, camera viewpoint, and inference latency. Continuous VLM inference during flight may introduce processing backlogs, increased hovering time, and additional propulsion energy consumption, directly reducing mission endurance. Consequently, semantic UAV inspection requires mission-aware scheduling strategies that jointly coordinate frame acquisition, inference triggering, and flight control under latency-energy constraints.

\textbf{Multimodal Training Data.}
Training semantic inspection models requires aligned multimodal datasets that combine infrastructure imagery with descriptive textual annotations. Unlike conventional detection datasets, semantic inspection demands contextual descriptions of structural conditions, defect severity, material degradation, and environmental context. However, publicly available multimodal infrastructure datasets remain extremely limited, particularly for UAV inspection scenarios. Generating high-quality semantic annotations is additionally costly and domain-specific, making data scarcity a major bottleneck for robust VLM-driven inspection systems.

\textbf{Generalization across Infrastructure Types.}
Infrastructure assets exhibit substantial variability in geometry, materials, degradation patterns, and environmental conditions. Defects such as cracks, corrosion, seepage, and displacement often appear differently across concrete, steel, and composite structures. Consequently, semantic inspection models must generalize across diverse infrastructure domains while maintaining reliable contextual reasoning. This requires structure-aware adaptation, domain-specific fine-tuning, and robust multimodal reasoning capabilities beyond fixed-category defect detection.

\textbf{Trustworthiness and Explainability.}
Infrastructure inspection often supports safety-critical decision-making, requiring that semantic outputs be reliable, interpretable, and verifiable. Unlike conventional object detectors, VLMs may generate hallucinated or inconsistent descriptions that are difficult to validate automatically. Ensuring trustworthy reasoning, explainable semantic outputs, confidence-aware reporting, and human-verifiable inspection findings is essential for practical deployment in UAV inspection systems.
%%%%%%%%%

\section{Feasibility Study: Edge-Native Semantic UAV Bridge Inspection}
\label{sec:IMPSetup}
%%%
\begin{table*}[!t]
\caption{Comparison of semantic inspections  by pre-trained and fine-tuned domain-adapted \texttt{SmolVLM-256.}}
\label{tab:exampleCommand}
\centering
    \begin{tabular}{| L{0.2cm} | L{4.4cm}  L{4.4cm} L{4.4cm}|}
    \toprule
        \rotatebox{90}{\textbf{Bridge Image Dataset}}
        &        
        \includegraphics[width=\linewidth]{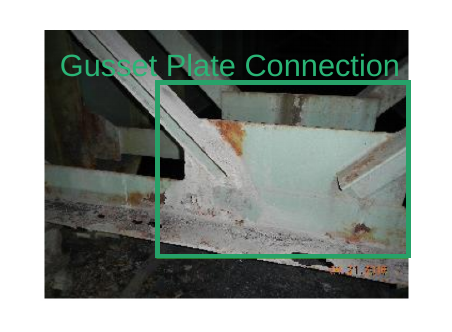} \newline
        \makebox[\linewidth][c]{\textbf{(a)}}  &         
        \includegraphics[width=\linewidth]{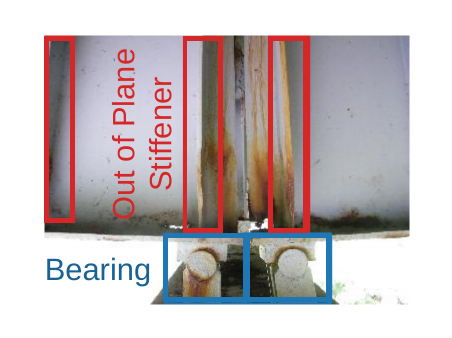} 
        \newline \makebox[\linewidth][c]{\textbf{(b)}}  &        
        \includegraphics[width=\linewidth]{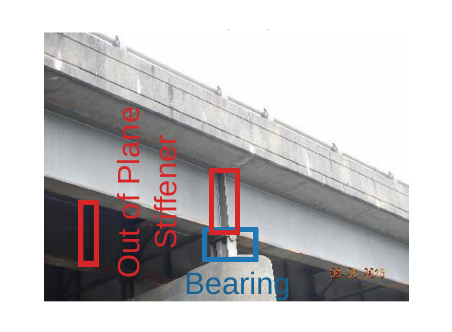} 
        \newline \makebox[\linewidth][c]{\textbf{(c)}} \\
        \midrule
        \rowcolor{gray!24}
        \rotatebox{90}{\textbf{Pre-trained}} &
        The surface of the iron is dirty and rusty. & 
        The rusty appearance of the components is consistent with a rusty surface. &  The concrete is gray and the metal is black. \\ 
        \midrule
       \rowcolor{cyan!15}
        \rotatebox{90}{\textbf{Fine-tuned}} &
        The components show signs of rust, as evidenced by the orange and brown discoloration and rough texture. There are also areas where the gray metal appears to be painted. & 
        % , but the rust is more concentrated
        The components' surfaces show significant staining and flaking, particularly at the edges and near the connections. The edges have orange and brown discoloration, indicating oxidation. & 

        The surfaces of the concrete and metal components appear to have a consistent texture, with no visible signs of discoloration or flaking. The concrete has a smooth finish with no visible cracks or pitting. \\
        \bottomrule
    \end{tabular}
\end{table*}
%%%%%%%%%%

\textbf{Evaluation Setup.}
To assess the feasibility of fully edge-based semantic inspection, we used an NVIDIA Jetson Orin representative of UAV-class edge hardware, equipped with an \num{8}-core CPU, \qty{16}{GB} RAM, \num{1024} CUDA cores, and \num{32} Tensor cores. We performed offline model fine-tuning on a server with a \num{128}{core} Intel Xeon Gold CPU and two NVIDIA Quadro GV100 GPUs. We employed the COCO-Bridge dataset~\cite{bianchi2021coco}, containing \num{774} bridge inspection images and more than \num{2500} annotated structural components, including bearings, gusset plate connections, cover plate terminations, and out-of-plane stiffeners.

For defect localization, we fine-tuned \texttt{YOLO26M}~\cite{yolo} using the Ultralytics framework with partial backbone freezing for \num{50} epochs at \mbox{$640\times640$} resolution using AdamW (\mbox{$5\times10^{-4}$}) and a batch size of \num{16}.
For semantic inspection, we adopted \texttt{SmolVLM-256}~\cite{marafioti2025smolvlm}, a lightweight \num{256}M-parameter VLM optimized using PEFT~\cite{han2024parameter} and LoRA~\cite{hu2022lora} applied to both the vision encoder and text decoder. We conducted fine-tuning via Hugging Face TRL (\texttt{SFTTrainer}) with fp16 precision (\mbox{$r=16$}, \mbox{$\alpha=32$}, dropout \num{0.15}, learning rate \mbox{$2\times10^{-4}$}, effective batch size \num{8}).
Since publicly available multimodal bridge inspection datasets remain limited, semantic captions describing structural surface conditions were automatically generated using \texttt{GPT-4o}.
%%%

\textbf{Component Detection vs. Semantic Inspection.}
As illustrated in Table~\ref{tab:exampleCommand}, \texttt{YOLO26M} accurately localizes bridge components such as gusset plate connections, bearings, and stiffeners using bounding-box detection, but provides limited information about structural condition. Conversely, the fine-tuned \texttt{SmolVLM-256} generates detailed inspection-oriented descriptions directly from captured imagery, identifying oxidation, staining, flaking, discoloration, and localized surface degradation patterns. Compared to the generic outputs produced by the pre-trained model, domain-specific fine-tuning substantially improves semantic inspection quality and contextual structural interpretation.
%%%%%
\begin{figure*}
\centering
  \subfigure[CPU, GPU, and RAM utilization.]{\includegraphics[width=0.4\textwidth]{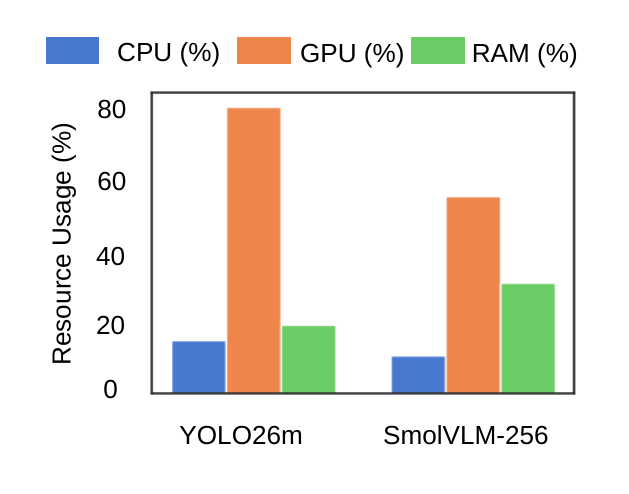}
  \label{fig:Fig3-a}}
  \subfigure[Inference latency and power consumption.]{\includegraphics[width=0.44\textwidth]{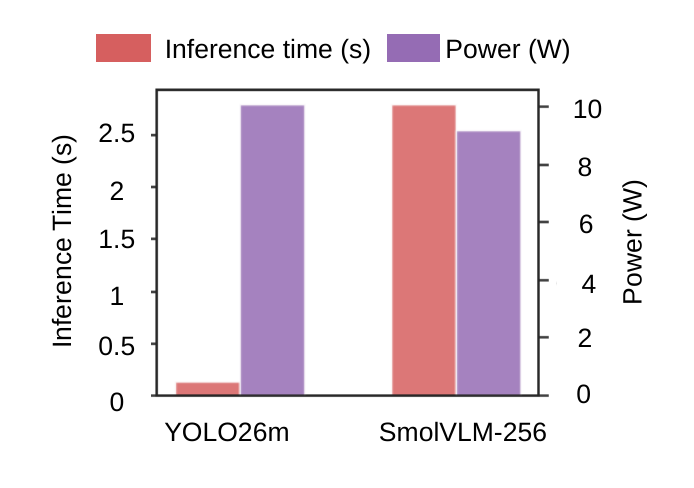}
  \label{fig:Fig3-b}}
  \subfigure[Temporal GPU utilization during inference.]{\includegraphics[width=0.42\textwidth]{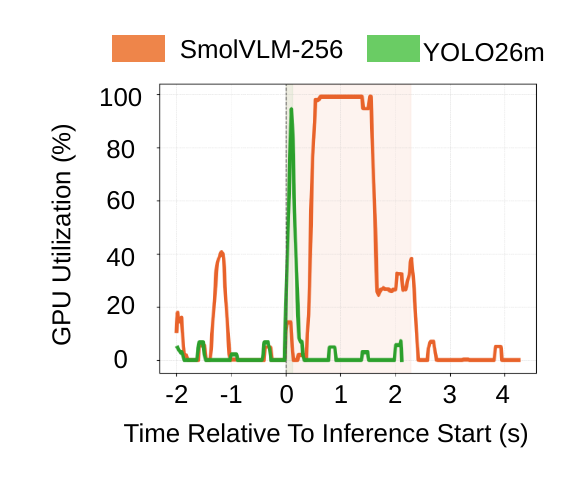}
  \label{fig:Fig3-c}}
  \subfigure[Temporal power consumption during inference.]{\includegraphics[width=0.42\textwidth]{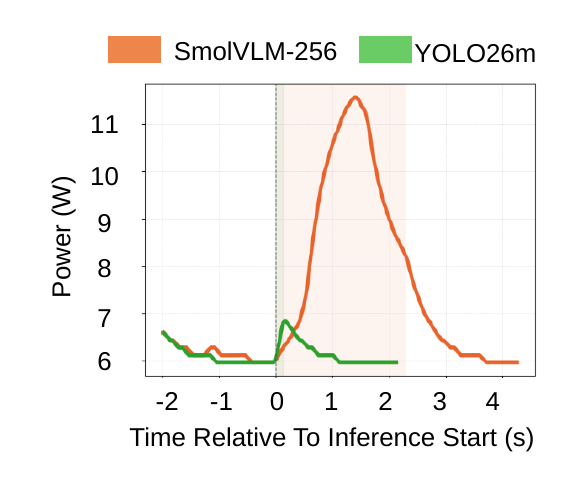}
  \label{fig:Fig3-d}}
\caption{Resource utilization and inference overhead of \texttt{YOLO26M} and \texttt{SmolVLM-256} on UAV edge hardware.}
\label{fig:resource}
\end{figure*}
%%%%%

\textbf{Edge Resource and Energy Analysis.}
To quantify operational overhead on UAV-class hardware, we monitored resource utilization using \texttt{tegrastats}~\cite{tegra} with \qty{1}{\milli\second} sampling intervals, recording CPU/GPU utilization, memory usage, inference latency, and total system power consumption. Computational energy was estimated over the inference interval and contextualized using the \texttt{DJI M210 V2} UAV battery profile (\qty{349.2}{Wh}, \num{24}-minute endurance). As shown in Fig.~\ref{fig:Fig3-b}, \texttt{YOLO26M} completes inference in approximately \qty{0.14}{\second}, while \texttt{SmolVLM-256} requires nearly \qty{2.8}{\second}. Although \texttt{YOLO26M} exhibits slightly higher instantaneous GPU utilization and peak power draw (Fig.~\ref{fig:Fig3-a}), its short execution window results in significantly lower cumulative energy consumption. 
In contrast, \texttt{SmolVLM-256} sustains elevated GPU and power utilization over a longer interval, primarily affecting inspection responsiveness while introducing additional computational energy overhead (Figs.~\ref{fig:Fig3-c}-\ref{fig:Fig3-d}). 

To evaluate cumulative operational impact, we further modeled inspection over a \qty{10}{\second}, \qty{30}{fps} video sequence (\qty{300} frames). \texttt{YOLO26M} processes the sequence in approximately \qty{42}{\second}, consuming only \qty{0.03}{\percent} of the UAV energy budget. In contrast, continuous \texttt{SmolVLM-256} inference requires nearly \qty{14}{\minute} while consuming approximately \qty{0.6}{\percent} of the battery capacity. Although computational energy remains modest relative to the UAV battery budget, inference latency is the dominant bottleneck, making continuous frame-by-frame VLM reasoning impractical. These results demonstrate that lightweight VLM-based semantic inspection is feasible on modern UAV hardware, but efficient deployment requires hierarchical pipelines in which lightweight detectors selectively trigger semantic reasoning only for candidate regions or key frames.

\section{Toward Autonomous Semantic UAV Inspection Systems} \label{sec:Discussion}
The results presented in this article indicate that lightweight VLMs can enable semantic infrastructure inspection directly on UAV-class edge hardware. However, the progression from component-level detection toward contextual structural understanding also introduces a broader transition in how future UAV inspection systems will operate across the edge-cloud continuum. Beyond isolated defect localization, next-generation inspection pipelines are expected to increasingly incorporate adaptive reasoning, collaborative intelligence, multimodal sensing, and context-aware understanding of infrastructure during flight.

\textbf{Agentic semantic inspection}, where lightweight VLMs are integrated with planning and decision-making modules to enable closed-loop inspection workflows. Instead of following static trajectories and continuously transmitting visual streams, future UAVs may dynamically adapt viewpoint selection, hover duration, sampling frequency, and inspection priorities based on semantic uncertainty and detected structural conditions. Such pipelines can reduce unnecessary sensing and communication overhead, improve inspection efficiency, and reduce energy consumption.

\textbf{Semantic digital twins}, where UAVs continuously update semantically enriched digital representations of bridges, tunnels, and power line assets using contextual structural observations collected during flight. Beyond isolated defect detection, such systems can support longitudinal degradation tracking, predictive maintenance, and condition-aware infrastructure management across edge-cloud environments.

\textbf{Collaborative edge intelligence}, where multiple UAVs jointly perform hierarchical sensing, cooperative perception, and distributed semantic reasoning across the computing continuum. Lightweight distilled models may execute locally on UAVs, while larger foundation models are selectively invoked at nearby edge servers or cloud infrastructure. Such architectures can enable scalable, large-area inspection while preserving real-time responsiveness and communication efficiency.

\textbf{Multimodal semantic inspection}, where visual reasoning is combined with complementary modalities such as thermal imaging, LiDAR, vibration sensing, acoustic analysis, and wireless sensing. Integrating sensing with semantic reasoning can improve defect interpretation, environmental awareness, and robustness under challenging operational conditions.

\textbf{Trustworthy and adaptive reasoning}, where future semantic inspection systems incorporate uncertainty estimation, retrieval-grounded reasoning, explainable outputs, and human-in-the-loop verification to improve reliability in safety-critical inspection tasks. In parallel, continual learning, synthetic semantic data generation, and adaptive multimodal fine-tuning will become increasingly important for supporting diverse infrastructure types and evolving degradation patterns. 
\section{Conclusion}
\label{sec:Conclusion}
This article presented the transition from cloud-centric visual processing toward fully edge-native semantic inspection powered by lightweight VLMs. It categorized existing UAV inspection architectures, discussed their associated system challenges and operational trade-offs, and demonstrated the feasibility of semantic edge intelligence on UAV-class hardware using fine-tuned YOLO26M localization and lightweight SmolVLM-256 reasoning. The results confirmed that semantic UAV inspection is technically feasible on modern UAV-class edge platforms, although scalable deployment requires hierarchical, resource-aware, and trustworthy semantic processing pipelines.

% \section*{Acknowledgment}
\bibliographystyle{IEEEtran}
\bibliography{bibtex}

\begin{IEEEbiographynophoto}{Reza Farahani}  
is a Univ.Ass. Postdoctoral Researcher at the Distributed Systems Group (DSG), TU Wien, and a Lecturer at the University of Klagenfurt, Austria, where he received his Ph.D. in Computer Science in 2023. He has recently coordinated the Austrian EdgeAI-Drone project and participated in the EU-funded ENFIELD Exchange Scheme on Green drone AI. Previously, he contributed to the Christian Doppler Laboratory ATHENA and EU Graph-Massivizer project, co-leading WP5 on serverless orchestration and large-scale edge-cloud testbeds. His research interests include distributed systems, serverless computing, distributed multimedia, agentic edge AI, and Green AI. Contact him at: r.farahani@dsg.tuwien.ac.at
\end{IEEEbiographynophoto}
%%%
\begin{IEEEbiographynophoto}{Naser Hossein Motlagh} is a Senior Researcher at the University of Helsinki, Finland. He received his D.Sc. in Networking Technology at Aalto University, Finland, in 2018. His research interests include the Internet of Things, wireless sensor networks, edge AI, environmental sensing, and unmanned aerial and underwater vehicles. Contact him at: naser.motlagh@helsinki.fi
\end{IEEEbiographynophoto}
%%%
\begin{IEEEbiographynophoto}{Zoha Azimi}  
is a Ph.D. candidate at the University of Klagenfurt, Austria. She received her M.Sc. in artificial intelligence from the University of Bologna, Italy. Her research interests include AI, energy-efficient multimedia systems, and agentic AI. Contact her at: zoha.azimi@aau.at
\end{IEEEbiographynophoto}
%%%
\begin{IEEEbiographynophoto}{Christian Timmerer}
is a Full Professor with the University of Klagenfurt, Austria, and the Director of the Christian Doppler Laboratory ATHENA. He is also a Co-Founder and Chief Innovation Officer of Bitmovin. His research interests include multimedia systems, adaptive streaming, and drone communication. Contact him at: christian.timmerer@aau.at
\end{IEEEbiographynophoto}
%%%
\begin{IEEEbiographynophoto}{Lorenzo Carnevale} is an Assistant Professor at the University of Messina, Italy. His research interests include edge intelligence and AI for resource-constrained environments. He has led technical activities in Horizon Europe projects on natural-disaster applications and distributed TinyAI. Contact him at lcarnevale@unime.it
\end{IEEEbiographynophoto}
%%%
\begin{IEEEbiographynophoto}{Sasu Tarkoma} is a Professor with the University of Helsinki and the University of Oulu, Finland. He received the Ph.D. degree in computer science from the University of Helsinki, in 2006. His research interests include mobile computing, data sciences, and AI. Contact him at: sasu.tarkoma@helsinki.fi
\end{IEEEbiographynophoto}
%%%
\begin{IEEEbiographynophoto}{Schahram Dustdar}  
is a Full Professor of Computer Science and Head of the Distributed Systems Group at TU Wien, Austria and ICREA Professor in Barcelona, Spain. He is serving as an Associate Editor for several IEEE and ACM journals and is EiC of Computing (Springer). His distinctions include the ACM Distinguished Scientist and Speaker and IBM Faculty Awards. He is member of the Academia Europaea. Contact him at: dustdar@dsg.tuwien.ac.at.
\end{IEEEbiographynophoto}
\end{document}